\documentclass[12pt]{iopart}
 
\usepackage{graphicx}   
\usepackage{xcolor}

\begin{document}

\title[]{Modelling efficient BB84 with applications for medium-range, terrestrial free-space QKD}
\author{Thomas Brougham and Daniel K. L. Oi}

\address{Computational Nonlinear and Quantum Optics, SUPA Department of Physics, University of Strathclyde, Glasgow G4 0NG, United Kingdom}
\ead{t.brougham@strath.ac.uk}
\vspace{10pt}
\begin{indented}
\item[]27 May 2022
\end{indented}

\begin{abstract}
Terrestrial free-space quantum key distribution is ideally suited for deployment in dense urban environments. The transition from laboratory to commercial deployment, however, raises a number of important engineering and deployment issues. Here, we investigate these issues for efficient BB84  using a weak coherent pulse-decoy state protocol.  We calculate expected key lengths for different environmental conditions and when the scope for optimisation of protocol parameters is restricted due to practical considerations. In particular, we find that for a fixed receiver basis choice probability, it can be advantageous to allow the transmitter to have a different basis choice probability depending on varying channel loss and background light levels.  Finally, we examine the effects of pulse intensity uncertainty finding that they can dramatically reduce the key length. These results can be used to determine the loss budget for the free-space optics of a QKD systems and assist in their design.
\end{abstract}


\section{Introduction}
\label{sec:intro}  

Quantum key distribution (QKD) allows for the secure distribution of cryptographic keys, where in principle, the security is guaranteed by the laws of physics~\cite{BB84,Scarani2009,wootters}.  This is in contrast to current public key methods, where the security is based on the difficulty of mathematical problems such as factoring or the discrete logarithm of elliptic curves~\cite{wootters,MathCryptography}.  Such schemes are rendered insecure by a quantum computer, which can efficiently solve the hidden abelian subgroup problem that underlies efficient solutions to factoring and discrete logarithms~\cite{Shor}. The development of QKD is thus vital to protect our communication infrastructure and provide forward security without reliance on computational complexity assumptions.  

In free-space (FS) QKD, signals are sent via line-of-sight transmission, e.g. between two ground stations or between a satellite and ground station.  Satellite-based FS-QKD has attracted a great deal of attention due to its potential to allow QKD over global distances\cite{ref1,ref2,ref3,ref4,ref5,ref6,ref7,ref8,ref9,Liao2017,Yin2020,Oi2017,Review}. In contrast, ground based FS-QKD typically operates over short to medium distances, ideal for deployment in dense urban spaces where it may be difficult and costly to deploy ad hoc short range fibre links. FS-QKD  between either buildings or existing towers should allow secure communication in a more flexible manner.
A prototype of such a network is being constructed as part of the AirQKD project~\cite{airqkdref}. Here, we present high-level analysis of terrestrial FS-QKD operational trade-offs with particular consideration of how finite key statistical effects and uncertainties impact upon their design.  

The development of a commercial ground based FS-QKD network faces significant challenges~\cite{Liorni2019}. Firstly, the system has to operate under diverse environmental conditions such as varying background light levels, meteorological visibilities~\footnote{Here, visibility refers to atmospheric transmittance~\cite{visibilityref}}, and levels of turbulence, etc.~\cite{weather2012}.  Furthermore, engineering or economic factors constrain the design and component choice. Components or operating settings may be standardised rather than optimised for each link scenario. For example, this could result in fixed or common protocol parameters such as basis bias and weak coherent pulse intensities set at the factory instead of being optimised for each deployment at greater cost. Hence, determining the performance of a fixed-setting FS-QKD system over a wide range of operating conditions and finding robust sets of protocol parameters are important practical considerations for real-world applications.

To ensure security, it is crucial to take into account finite statistical fluctuations and uncertainties during operation. A vital step for all QKD protocols is to use the transmitted data to estimate parameters, e.g. such as the quantum bit error (QBER), yields from vacuum and single photon emissions, and phase errors~\cite{decoy2,decoy3,decoy4}. In practical use, the transmission window (or {\it integration time}) is finite and will be predetermined or limited due to operational considerations such as key refresh tempo or latency. Finite statistical effects can lead to sharp cutoffs in the positive secret key region as a function of channel loss and background count, thus selection of block accumulation lengths and integration times should be balanced against operational requirements.

QKD protocols based on weak coherent pulses (WCPs) requires knowledge of their intensities and these must be stabilized to what is intended.  This is achieved by using power monitors to monitor the power and adjusting~\cite{Lucamarini2013}.  However, for any measurement there is an uncertainty associated with this.  In a laboratory setting this uncertainty can be reduced and controlled.  But in real world deployment, we must expect there to be a non-negligible uncertainty.  If we are to ensure security, we must increase the amount of privacy amplification to take account of this uncertainty. We study this effect and show that it leads to both a reduction of the secure key length and a reduction of the region of parameter space for which a secret key can be extracted.  

In this paper, we perform a systematic study of efficient BB84 across different environmental conditions, for fixed protocol parameters and study the effects of uncertainties in the intensities. In principle, the results can also be applied to fibre QKD, though will be of lesser utility as their operating conditions tend to be more stable compared with a terrestrial FS-QKD link that will be subject to large variations in loss and background light throughout the day. Geographic variation between FS-QKD links also requires modelling of the performance for a wide range of conditions and typical operating points.  The outline of the manuscript is as follows.  In section~\ref{sec:channel} we describe the channel and QKD setup; which serves to introduce the main physical parameters.  In section~\ref{sec:protocol} we explain the protocol and how this can be implemented physically in a real system.  Results for optimizing the length of the secret key under different conditions are presented in section~\ref{sec:results}.  Next, we study the effects of fixing protocol parameters on the optimization of the secure key length in section~\ref{sec:fixed}.  We then investigate the effects of uncertainty in the WCP intensities in section~\ref{sec:uncertainty}.  Finally, we discuss the findings in section~\ref{sec:conc}.  Preliminary versions of some of these result can also be found in \cite{spiepaper}.

\section{The free-space system}
\label{sec:channel}

We consider a free-space optical line-of-sight system between a transmitter (Alice) and receiver (Bob) with a 100MHz pulsed source of phase randomized polarisation encoded WCPs with signals states: vertical ($0^\circ$), horizontal ($90^\circ$), diagonal ($45^\circ$) and anti-diagonal ($135^\circ$), $V$, $H$, $D$ and $A$, respectively. For the purposes of a high-level system performance model, we assume that side-channel leakage is negligible~\cite{sidechannel}. In practice, any implementation should be evaluated for such vulnerabilities, non-idealities characterised, and their effects taken into account in the final security analysis.

In a deployed system the maximum integration time will normally be preset and cannot be easily increased without increased memory or processing. It is thus vital to evaluate system performance over a range of integration times, or transmission block sizes that are the product of the pulse repetition rate and the integration time. 
The secure key length is determined by the total number of pulses transmitted and received, not the transmission rate, hence we consider a fixed repetition rate with different integration times to determine the number of pulses sent. Results for different repetition rates and transmission times can be easily inferred, keeping other parameters fixed.

Free-space channels suffer losses due to beam spreading (i.e. diffraction or geometrical losses), absorption, and scattering~\cite{fso2010}. Turbulence will also cause the beam to both wander and distort~\cite{turb2002,turb2016}, reducing the coupling of the modes to the collection device. Sway of the transmitters and receivers together with additional pointing errors also adds to loss. These effects could be mitigated by active correction methods \cite{ref10}. Other source of loss are imperfect mode coupling into the detector, detector efficiencies, and losses within the transmitter and receivers optics.

The losses in the transmitter can be mitigated, provided they can be well characterized. For security, what matters is that the emitted signal intensities exiting the transmitter aperture and into the Eve-controlled quantum channel are as specified. We can compensate for internal transmitter losses by producing WCPs with greater intensities at the source so that when they leave the exit aperture, they have the required intensity. Scattered light within the transmitter needs to be prevented from escaping else it could leak side-channel information.

We assume that transmitter losses are compensated by the above method hence we only need to consider the remaining contribution that we group together to obtain a total system loss $\eta_{\mathrm{loss}}$.  Let $p_{d}$ be the total probability that a photon emitted from the transmitter is detected at the receiver, we characterise the channel optoelectronic efficiency by the total system loss (in dB) defined as
\begin{equation}
\eta_{\mathrm{loss}}=-10\log_{10}(p_d).
\end{equation}

We also consider {\it extraneous counts} due to detector dark counts or from stray-light where $p_{ec}$ is the probability per pulse. Another source of errors is due to errors in the polarization states or their measurement and we characterise these by the {\it intrinsic quantum bit error rate} or $QBER_I$ that is independent of channel loss.  We also account for detector after-pulsing where we assume the value $p_{ap}=10^{-3}$, consistent with values used in previous modelling of fibre and free-space QKD systems~\cite{Lim2014,Sidhu2021}. For convenience, fixed modelling parameter are listed in table \ref{tab:constant}.


\begin{table}[ht]
\caption{A table of parameters that we keep constant throughout modeling.} 
\label{tab:constant}
\begin{center}       
\begin{tabular}{|l|l|l|}
\hline
\rule[-1ex]{0pt}{3.5ex}  After-pulse probability & $p_{ap}$ & $1\times 10^{-3}$\\
\hline
\rule[-1ex]{0pt}{3.5ex}  Source's repetition rate & $f_s$ & $100$MHz\\
\hline
\rule[-1ex]{0pt}{3.5ex}  Correctness parameter & $\epsilon_{c}$ & $10^{-15}$\\
\hline
\rule[-1ex]{0pt}{3.5ex}  Secrecy parameter & $\epsilon_{s}$ & $10^{-9}$ \\
\hline
\end{tabular}
\end{center}
\end{table}

The photon statistics of the WCP signal states are described by a Poissonian distribution $P(n)=\exp(-\mu)\mu^n/n!$ where $\mu$ is the mean number of photons per pulse. An eavesdropper can perform a photon number splitting attack on the multiphoton emissisions~\cite{decoy1,pnsattack}. To counter this, we use the so-called decoy state method to estimate the number of received vacuum and single photon emission events~\cite{decoy2,decoy3,decoy4}.

\section{The QKD protocol}
\label{sec:protocol}
\subsection{Outline of protocol}
We employ the Efficient BB84 WCP Decoy State protocol with 3 intensities $\mu_1$, $\mu_2$ and $\mu_3$ that are randomly and independently chosen~\cite{Lim2014}~\footnote{In other literature, this is also known as the 2-decoy state protocol.}. Alice prepares WCPs with polarization chosen from one of two bases: $X\in\{H,V\}$ and $Z\in\{D, A\}$.  With probability $P^A_X$ Alice chooses the $X$ basis, and with probability $P^A_Z=1-P^A_X$ she chooses the $Z$ basis,  In efficient BB84 the two bases need not be chosen with equal probability.  Instead, the probability is found by optimizing the protocol so as to maximize the secure key length.  With the basis chosen, Alice randomly and uniformly encodes bit values 0 or 1 onto the appropriate polarization state, where: $V,D\equiv 0$ and $H,A\equiv 1$, before choosing the intensity of each pulse: $k\in\{\mu_1,\mu_2,\mu_3\}$ with probability $p_k$. The intensity values can be any values that satisfy the relations: (i) $\mu_1>\mu_2>\mu_3$ and (ii) $\mu_1>\mu_2+\mu_3$.  Bob randomly chooses to measure the polarization in either the basis $X$ or $Z$, with probability $P^B_X$ and $P^B_Z=1-P^B_X$, respectively.  

After transmission Alice and Bob perform basis sifting through authenticated public announcement where they retain only those results corresponding to received pulses when they chose the same basis. Alice and Bob will then have $n_X$ sifted $X$ basis bits and $n_Z$ sifted $Z$ basis.

For the sifted bits, Alice publicly announces her choice of intensities for each WCP allowing Bob to tag each bit with its transmitted intensity. Let $n_{X(Z),k}$ denote the number of bits in the $X(Z)$ basis that originated from a pulse of intensity $k$.  The secret key is constructed using only the sifted bits from the $X$ basis, while the bits from the $Z$ basis are used for parameter estimation.  All the data from the $Z$ basis is publicly announced, which allows Alice and Bob to calculate the number of errors in this basis: $m_{Z}$.  Additionally, they can calculate the number of errors in bits that originate from pulses with intensity $k$, $m_{Z,k}$.  Using these results, one can calculate the phase error $\phi_X$ using Eq.s (4) and (5) of~\cite{Lim2014}.  

Alice and Bob's sifted bit string for the $X$ basis will differ due to errors, which are corrected in the reconciliation stage, to produce the identical bit strings called the {\it raw key}.  Error correction can be performed using a number of schemes~\cite{wootters}, all of which requires public communication between that leaks $\lambda_{EC}$ bits.  Alice and Bob then perform a verification step to ensure their raw keys are identical with probability at least $1-\epsilon_{c}$, where $\epsilon_{c}$ is a pre-agreed correction parameter~\cite{Lim2014}, here chosen to be $\epsilon_{c}=10^{-15}$.  The final stage is privacy amplification~\cite{Scarani2009,wootters,rennerthesis} that reduces the key to length $\ell$ bits.

\subsection{Secure key length}

From the analysis of~\cite{Lim2014}, the final composable secure key length is given by
\begin{equation}
\label{skl}
    \ell  = \Big\lfloor s_{X,0}  + s_{X,1} (1 - h(\phi_{X}))
     - \lambda_{EC} - 6 \log_2 \frac{21}{\epsilon_{s}} - \log_2 \frac{2}{\epsilon_{c}}\Big\rfloor,
\end{equation}
where $s_{X,0}$ is the estimated number of bits coming from vacuum events, $s_{X,1}$ is the number of bits originating from single photons, $\epsilon_{s}$ is the security parameter  and $h(x)=-x\log_2 (x)-(1-x)\log_2 (1-x)$ is the binary entropy. A protocol is said to be $\varepsilon=\epsilon_s +\epsilon_c$ secure if it is $\epsilon_c$-correct and $\epsilon_s$-secret~\cite{rennerthesis}. For all calculations presented we use $\epsilon_c =10^{-15}$ and $\epsilon_s =10^{-9}$.

The quantities $s_{X,0}$, $s_{X,1}$ and $\phi_X$ can be calculated using Eq. (1) to (5) from~\cite{Lim2014}.  In that paper, there are correction terms due to finite statistical fluctuations appearing in the quantities $n^{\pm}_{X(Z),k}$ and $m^{\pm}_{Z,k}$, which we define below in Eq. (\ref{finitecor}). In~\cite{Lim2014} the correction terms are calculated using the Hoeffding bound~\cite{hoeff63} but it is possible to improve them using a modified version of the Chernoff bound~\cite{Zhang2017_PRA,Yin2020_SR} as described in~\cite{Sidhu2021}. For the current analysis we use
\begin{eqnarray}
\label{finitecor}
n^{\pm}_{X(Z),k}&=\frac{e^k}{p_k}\left[n_{X(Z),k}\pm \delta^{\pm}_{n_{X(Z),k}} \right],\\
m^{\pm}_{X(Z),k}&=\frac{e^k}{p_k}\left[m_{X(Z),k}\pm \delta^{\pm}_{m_{X(Z),k}} \right],
\end{eqnarray}
where $k\in\{\mu_1,\mu_2,\mu_3\}$ and 
\begin{eqnarray}
\label{finitecor2}
\delta^{+}_{Y}=\beta+\sqrt{2\beta Y +\beta^2}, \quad
\delta^{-}_{Y}=\frac{\beta}{2}+\sqrt{2\beta Y +\frac{\beta^2}{4}},
\end{eqnarray}
where $\beta=\ln(1/\varepsilon)$ and $Y$ can equal either $n_{X(Z),k}$ or $m_{X(Z),k}$~\cite{Yin2020_SR}.

The environmental parameters $\eta_{\mathrm{loss}}$, $p_{ec}$ and $QBER_I$ appear within the expressions for $n_{X(Z),k}$ and $m_{X(Z),k}$, which feed into Eqs. (\ref{finitecor}) and (\ref{finitecor2}).  For more details on how the environmental factors are incorporated within the model, see appendix E or \cite{Sidhu2021}.

The error correction term $\lambda_{EC}$ is equal to the classical information exchanged during the reconciliation stage, this is included in the estimate of the amount of privacy amplification needed.  In practice, we can keep track of the bits exchanged and thus $\lambda_{EC}$ is known and not subject to statistical uncertainty.  However, to model the systems performance we need an estimate of $\lambda_{EC}$ and we use the sophisticated estimate introduce in~\cite{Tomamichel2017}, as discussed in~\cite{Sidhu2021}.

\subsection{Unequal basis choice}
\label{sec:unequal}
In Efficient BB84, it is standard to choose $P^A_X=P^B_X$ and optimised for the channel conditions~\cite{Lo2005}.  In practice, however, it often straightforward for Alice to vary her basis probabilities but difficult for Bob.   This is because Alice can encode polarization states using four separate WCP emitters. Changing $P^A_X$ amounts to changing the probability of Alice choosing the X basis signals.  In contrast, Bob's measurement is typically realized by beam-splitter that randomly sends photons to one of two polarization measurement setups, to measure either in the basis $X$ or $Z$.  While a variable beam-splitter could be used in a laboratory, it would be more economical and practical in a mass produced commercial system to use a fixed beam-splitter.

To understand why setting $P^A_X\ne P^B_X$ is helpful in the case of Bob's receiver being fixed, consider the sifting stage.  Before sifting, Alice and Bob shared $N$ bits.  
The fraction of these bits discarded in the sifting stage is $P^A_X P^B_Z+P^A_Z P^B_X$. The fraction of bits retained in the $X$ basis is $P^A_X P^B_X$, while $P^A_Z P^B_Z$ is the fraction in the $Z$ basis.  To maximize the key length, we want $P^A_X P^B_X$ close to one, while still allowing enough bits in the $Z$ basis to accurately estimate the parameters.  If $P^B_X$ is fixed and $P^A_X=P^B_X$,  we have no control of the sifting ratios.  
We thus generalize the protocol by allowing Alice and Bob to have different basis choice probabilities.  This change in the protocol affects the numbers $n_{X(Z),k}$ and $m_{X(Z),k}$ seen in a given run.  The change, however, does {\it not} affect either the key length formula, (\ref{skl}), or crucially, the security analysis~\cite{Lim2014,rennerthesis}.  

\subsection{Optimization of the protocol}
\label{sec:optimize}
For a given value of the integration time, source repetition rate, $\eta_{\mathrm{loss}}$, $QBER_I$ and $p_{ec}$, we will calculate the secure key length.  Initially, this involves optimizing the secure key length by varying the parameters: $\{P^A_X, p_{\mu_1},p_{\mu_2},\mu_1,\mu_2\}$.  The value of $P^B_X$ will either be set equal to $P^A_X$ or will be fixed.  The smallest intensity, $\mu_3$ should be set to zero, but to avoid numerical issues, we instead use a very small value, $\mu_3=10^{-9}$.  The details of the optimization are the same as those presented in~\cite{Sidhu2021} with the exception that now the channel loss is assumed to be fixed in time.  

In section \ref{sec:fixed} we fix $P^B_X$ and the intensities $\{\mu_1, \mu_2,\mu_3\}$.  The numerical optimization of the secure key length now involves varying only the parameters: $\{P^A_X, p_{\mu_1},p_{\mu_2}\}$, this only requires changing the way that the unbiased random number stream, e.g. from a quantum random number generator (QRNG), are processed to derive the source input data.  The approach is the same as before, but where now we input fixed values for $P^B_X$, $\mu_1$, $\mu_2$ and $\mu_3$ which we assume are set and calibrated at the factory. The values of $P^B_X$ chosen correspond to splitting rations of commercially available beam-splitters.  The fixed values for the intensities are picked by a process of trial and error as well as through studying typical values found in the cases where they were optimized.  The aim was not to find the optimal choice of fixed intensities, but instead to show that good choices existed over a range of channel conditions.  As such the values of fixed values of intensities described in section \ref{sec:fixed} were not numerically optimized, but instead found using a process of trial and error.

\section{System performance}
\label{sec:results}

For the design of free-space optical systems, one must know the total loss budget, i.e. the maximum $\eta_{\mathrm{loss}}$ one can tolerate, while still producing a secure key. 
We evaluate the secure key length for different values of $\eta_{\mathrm{loss}}$, $p_{ec}$ and $QBER_I$. The results we present allow one to both evaluate the system performance under a wide range of environmental conditions and to determine a total loss budget for a given set of circumstances.

\begin{figure} [ht]
\centering
\includegraphics[width=0.7\textwidth]{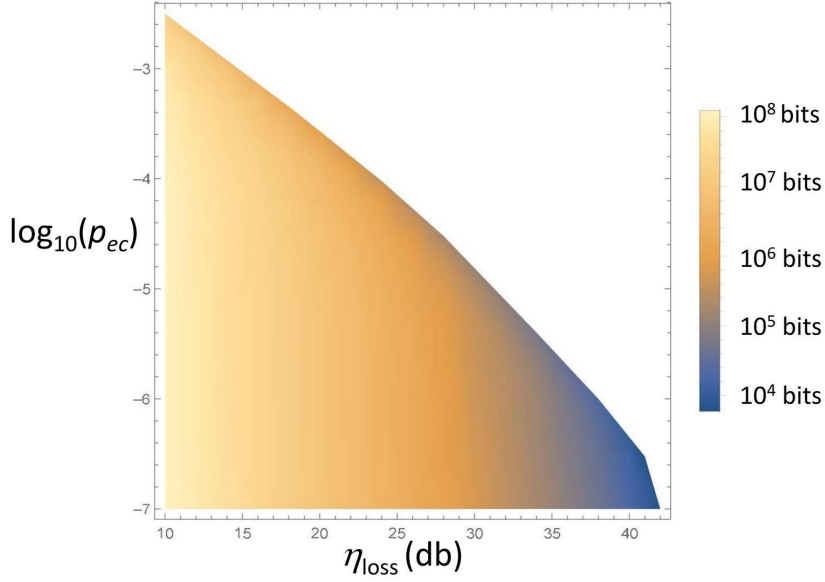}
\caption[example] 
   { \label{fig:heatmap1} 
A 3D plot showing values of the secret key for different values of the total system loss, $\eta_{\mathrm{loss}}$, and the $\log_{10}(p_{ec})$.  The plot is for $QBER_I=0.005$ and an integration time of 60 seconds.}
   \end{figure}

\begin{figure} [ht]
    \centering
    \includegraphics[height=5cm]{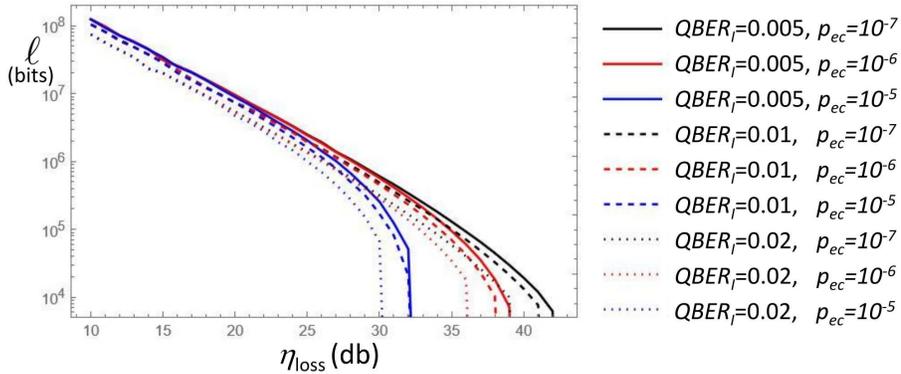}
    \caption[example]
    {\label{fig:diffQBER}
A plot of $\ell$, the secure key length (bits), against $\eta_{\mathrm{loss}}$, the total system loss.  We plot for different values of $QBER_I$ and different values of $p_{ec}$.  The solid lines correspond to $QBER_I=0.005$, the dashed lines are for $QBER_I=0.01$, while the dotted lines are for $QBER_I=0.02$.  The Black curves are for $p_{ec}=10^{-7}$, the red curves are for $p_{ec}=10^{-6}$, while the blue curves are for $p_{ec}=10^{-5}$.} 
\end{figure}

Consider a system operating with an integration time of 60s (link refresh period), $QBER_I=0.5\%$, $P^A_X=P^B_X$ and repetition rate of 100MHz. A plot showing the region where one obtains secret keys is shown in figure \ref{fig:heatmap1} where we have numerical optimised all protocol parameters at each point, as discussed in section \ref{sec:optimize}. The shaded region shows where one obtains a nonzero value for the secret key. We observe that $\ell$, when non-zero, is large, even for relatively large losses.  
An important feature of the plot is that a slight increase in $p_{ec}$ or $\eta_{\mathrm{loss}}$ can cause a sharp drop off in the secret key or reduce it to zero.  This is illustrated more clearly in figure \ref{fig:diffQBER} where we plot $\ell$ against $\eta_{\mathrm{loss}}$, for different values of $QBER_I$ and $p_{ec}$. An important feature is the sharp drop of $\ell$ to zero resulting from finite statistical corrections \cite{Sidhu2021}.


From figure \ref{fig:heatmap1} we observe that the effect of increasing $p_{ec}$ becomes more pronounced as $\eta_{\mathrm{loss}}$ increases, this is due to two reasons.  Firstly, increasing the loss decreases the total counts and thus increasing finite statistical fluctuations in the parameters.  Errors due to $p_{ec}$ thus become more significant.  Secondly, as we increase $\eta_{\mathrm{loss}}$, the fraction of the total counts corresponding to extraneous counts increases.  This increases both the quantum bit error rate and the phase error, $\phi_X$. Figure \ref{fig:heatmap1} strongly suggest that $p_{ec}$ greatly impacts the secure key length.



It is also important to understand the effect of the intrinsic quantum bit error rate, $QBER_I$ on $\ell$.  In figure \ref{fig:diffQBER} we see that increasing $QBER_I$ causes $\ell$ to decrease.  Furthermore, increasing both $QBER_I$ and $p_{ec}$ causes $\ell$ to drop to zero for smaller values of the total loss.  We see however, that increasing $QBER_I$ from 0.005 to 0.01 does not affect $\ell$ as greatly as increasing $p_{ec}$.   

\begin{figure} [ht]
\begin{center}
\begin{tabular}{c} 
\includegraphics[height=7cm]{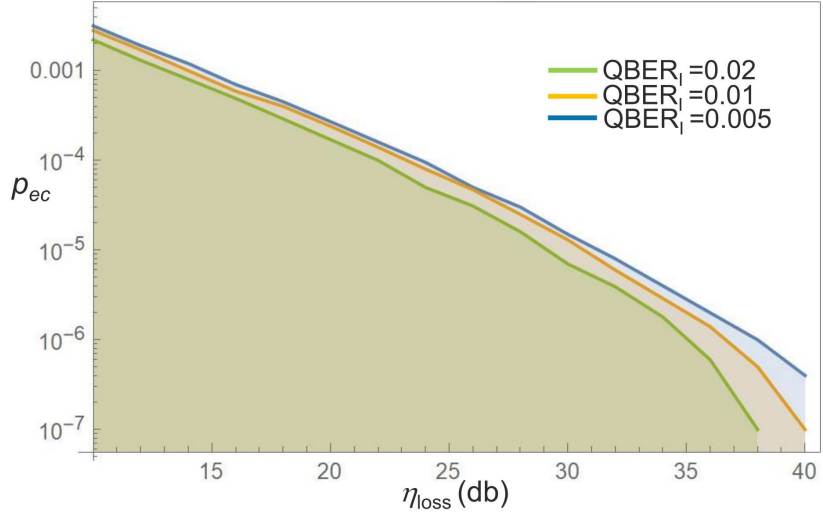}
\end{tabular}
\end{center}
   \caption[example] 
   { \label{fig:region} 
A plot showing the region where $\ell>0$, i.e. when we can extract a secure key}, for different values of $QBER_I$. All curves are for an integration time of 60 seconds.  The coloured lines represent the boundary for $\ell>0$, while the shaded coloured regions also corresponds to $\ell>0$.
   \end{figure}

The sharp drop off of $\ell$ to zero is important when designing QKD systems.  Is vital to know the range of parameters for which $\ell >0$.  In figure  \ref{fig:region} we plot the region of values for $\eta_{\mathrm{loss}}$ and $p_{ec}$ where $\ell>0$.  The regions are plotted for different values of $QBER_I$.  It is evident that increasing $QBER_I$ not only decreases $\ell$, as shown in figure \ref{fig:diffQBER}, it also leads to a decrease in the regions where one can extract a secret key.  Notice, however, for moderate values of $\eta_{\mathrm{loss}}$ the difference in regions for values of $QBER_I$ equal to $0.5\%$ and $1\%$ is small.  This suggests that if the QKD network is expected to operate in conditions where $\eta_{\mathrm{loss}}$ is below 35 dB, then the effort to decrease $QBER_I$ below $1\%$ might not yield great rewards.  Instead, more attention should be focused on reducing $p_{ec}$.


\begin{figure} [ht]
\begin{center}
\begin{tabular}{c} 
\includegraphics[height=6cm]{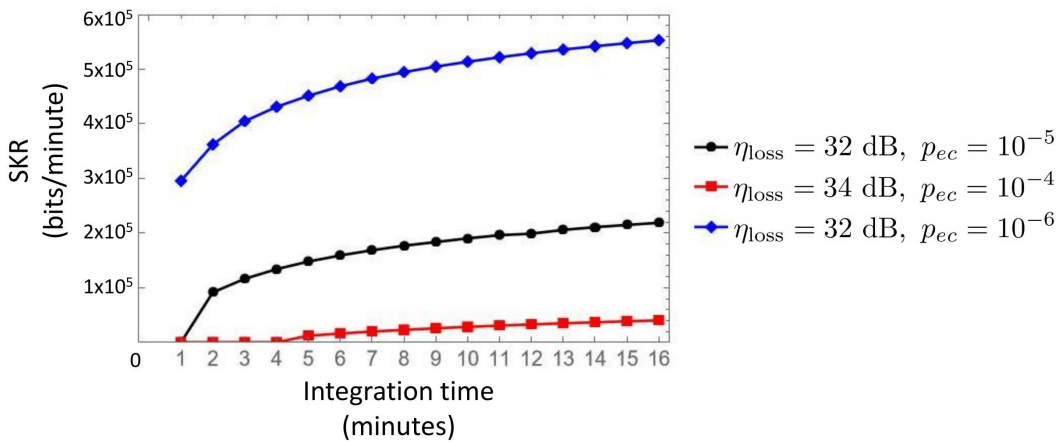}
\end{tabular}
\end{center}
   \caption[example] 
   { \label{fig:time} 
A plot showing the secret key rate (SKR), i.e. secret bits per minute, against the integration time.  The blue curve is for $\eta_{\mathrm{loss}}=32$ dB and $p_{ec}=10^{-6}$ , the black curve is for $\eta_{\mathrm{loss}}=32$ dB and $p_{ec}=10^{-5}$ and the red curve is for $\eta_{\mathrm{loss}}=34$ dB and $p_{ec}=10^{-4}$.  All three curves where generated with $QBER_I=0.005$. }
   \end{figure}

The integration time also affects the secret key rate (SKR).  Doubling the integration time doubles the raw key bits, this may be expected to also double $\ell$ but several effects can lead to a super-linear increase. A greater integration time also yields more data for parameter estimation, hence we can estimate the parameters with less error and obtain a smaller correction due to the finite statistics, illustrated in figure \ref{fig:time}. In this plot, all curves correspond to $QBER_I=0.005$.  Both the blue and black curves are for a total system loss of $\eta_{\mathrm{loss}}=32$ dB.  The extraneous count probabilities are $p_{ec}=10^{-6}$ for the blue curve and $p_{ec}=10^{-5}$ for the black curve.  In contrast, the red curve is for $p_{ec}=10^{-5}$ and $\eta_{\mathrm{loss}}=34$ dB. If increasing the integration time had no effect on parameter estimation, then the SKR would be constant as integration time increases.   However, as we see in figure \ref{fig:time}, the SKR increases with integration time.  In both the black and red curves, $\ell=0$ initially.   However, by increasing the integration time we eventually obtain a nonzero secret key.  Furthermore, in all three curves, while the SKR continues to increase as the integration time increases, the rate of increase slows, suggesting that it will plateau for a sufficiently large integration times.  These results dramatically illustrate the effects of finite statistics.  For integration time of 60 seconds, the errors within parameter estimation could lead to $\ell=0$.  However, by increasing the integration time we can extract a secret key.  

These results show that if possible, one should avoid using several short transmission windows.  Instead, one should use fewer, but longer transmission windows.  This would necessitate increasing the systems memory and changing the key generation cycle to accommodate the production of large amounts of key, over longer intervals.  In principle, one could also group several smaller transmission windows together to form a larger integration time. This approach was investigated in~\cite{Sidhu2021} in the case of limited satellite overpass time, but here could be used to mitigate against weather induced channel outages. This is valid for finite data security proofs that use smoothed min-entropies, as described in \cite{Lim2014}.

\begin{figure*} [ht]
\begin{center}
\begin{tabular}{c} 
\includegraphics[width=\textwidth]{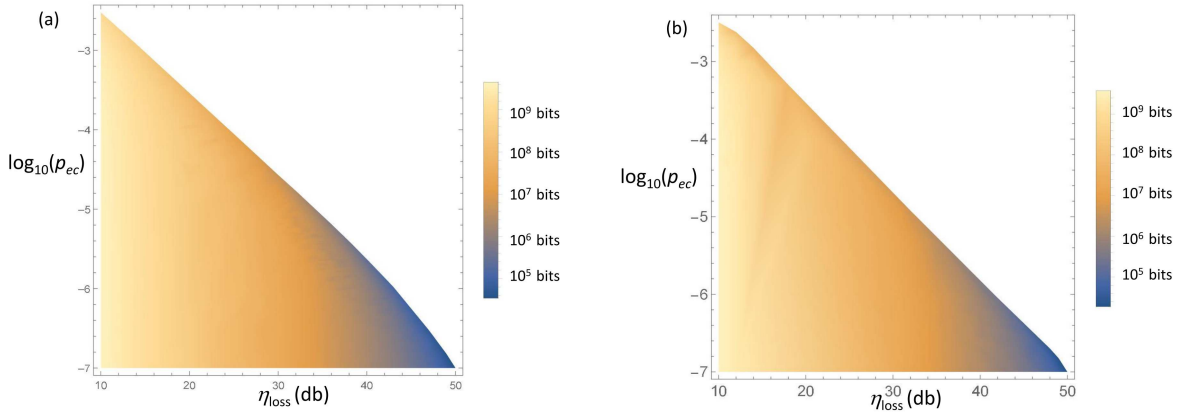}
\end{tabular}
\end{center}
   \caption[example] 
   { \label{fig:min30} 
A 3D plot showing values of the secret key for different values of the total system loss, $\eta_{\mathrm{loss}}$, and the $\log_{10}(p_{ec})$, for an integration time of 30 minutes.  Plot (a) corresponds to $QBER_I=0.005$, while (b) is for $QBER_I=0.01$.  Notice that for the long integration time of 30 minutes, plots (a) and (b) are very similar. This shows that doubling $QBER_I$ from 0.005 to 0.01 does not have a strong effect.} 
\end{figure*}

In view of the previous results, one should use large integration times.  For this reason we study the secure key length with an integration time of 30 minutes.  Under the correct circumstances, this will provide a large amount of secret key.  One can then use this to distribute multiple secret keys within a QKD network.  Figure \ref{fig:min30} shows a plot of the secret key against $\eta_{\mathrm{loss}}$ and $p_{ec}$, where (a) is for $QBER_I$ equal to $0.5\%$ and (b) is for $QBER_I$ equal to $1\%$.  The long integration times means that we can extract secret keys for large losses. For example, for $QBER_I=0.005$, we can extract 34,256 bits for $\eta_{\mathrm{loss}}=50$ dB and $p_{ec}=10^{-7}$.  The plots also show the importance of background light.  For example, with $p_{ec}=10^{-3}$ we can only extract a secret key for $\eta_{\mathrm{loss}}$ up to 14 dB. An important feature of figure \ref{fig:min30} is that plots (a) and (b) are very similar.  This demonstrates that increasing $QBER_I$ from $0.005$ to $0.01$ has an almost negligible effect for large integration times. This suggests that once $QBER_I$ has fallen ~0.01, there are diminishing returns from its further improvement. For examples of how these results can be used to obtain loss budgets, see appendix A.

\section{Performance with fixed parameters}
\label{sec:fixed}
In a real world deployment of a QKD system, it may not be cost effective to produce separate systems with different parameters for different operating conditions due to the additional effort and components required.  
An obvious issue is that Bob's basis choice is made using a fixed beam splitter.  This means that $P^B_X$ will be fixed, unlike in the previous plots where we assumed one could choose any value.  In contrast, as we explained in section \ref{sec:unequal}, Alice's can easily vary her basis choice probability, $P^A_X$.  In the remainder of the analysis we allow for $P^A_X \ne P^B_X$.  We consider a long integration time of 30 minutes and $QBER_I$ of $1\%$.  In models of fibre and satellite QKD systems, a value of $0.5\%$ has often been assumed~\cite{Lim2014,Sidhu2021}.  The current value for $QBER_I$ is more conservative and reflects the fact that a mass produced commercial system may not be as well aligned as current experimental systems.

\begin{figure} 
\begin{center}
\begin{tabular}{c} 
\includegraphics[width=\textwidth]{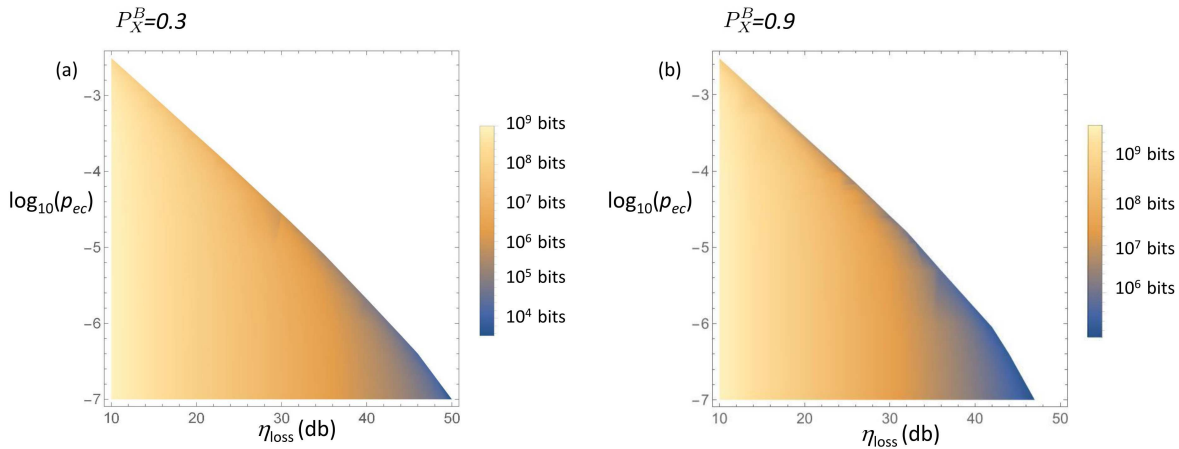}
\end{tabular}
\end{center}
   \caption[example] 
   { \label{fig:fixed1} 
A 3D plot of$\ell$ plotted for total system loss $\eta_{\mathrm{loss}}$ and $\log_{10}(p_{ec})$.  All plots are for $QBER_I=0.01$ and an integration time of 30 minutes.  The receiver basis choice probability is fixed to set values for both plots.  In figure (a) we fix $P^B_X=0.3$, while in (b) we have $P^B_X=0.9$.  This figure illustrates that there can be a trade-off for the choice of $P^B_X$.  Low values for $P^B_X$ give $\ell>0$ for wider ranges of losses than larger values of $P^B_X$.  However, when a secure key is produced, it tends to be larger when $P^B_X$ is closer to one.}
   \end{figure} 

In figure \ref{fig:fixed1} we present plots of the optimized values for $\ell$, where $P^B_X$ is fixed to be a particular value.  The optimization is performed by varying the values of $\{P^A_X, p_{\mu_1},p_{\mu_2},\mu_1,\mu_2\}$.  In figure \ref{fig:fixed1}, (a) is for $P^B_X=0.3$, while (b) is for $P^B_X=0.9$.  Additional plots for $P^B_X=0.5$ and $P^B_X=0.7$ are shown in figure \ref{fig:D_fig1} of appendix B. One can extract a secret key over a larger range of $\eta_{\mathrm{loss}}$ for smaller values of $P^B_X$.  For instance, for $P^B_X=0.9$, the largest total system loss for which we can extract a secret key for is $\eta_{\mathrm{loss}}=47$ dB.  In contrast, we can extract secret key up to $\eta_{\mathrm{loss}}=50$ dB for both $P^B_X=0.5$ and $0.3$.  While we can extract a secret key for a wide range of parameter, when we fix $P^B_X$, figure \ref{fig:fixed1} shows that there is a trade-off for different values of $P^B_X$.  For $P^B_X=0.9$ we obtain a larger value of $\ell$ at smaller losses, than we do for $P^B_X=0.3$.  For example, for $\eta_{\mathrm{loss}}=10$ dB and $p_{ec}=10^{-3}$ we obtain $2.42\times 10^9$ bits when $P^B_X=0.9$ and $8.68\times 10^8$ bits for $P^B_X=0.3$. Using $P^B_X=0.9$ yields a key that is $\approx 2.8$ times greater than using $P^B_X=0.3$.  However, fixing $P^B_X$ to a small value gives $\ell>0$ over a wider range of losses, but at a cost of providing less secret key bits at lower losses than using larger values for $P^B_X$.  This trade-off results as lower values for $P^B_X$ means $P^B_Z$ is larger thus increasing the number of bits available for parameter estimation. This is important for high losses, but for lower losses this leads to an excess of signals for parameter estimation at the expense of key generation.  As only bits in the $X$-basis are used to construct the final key, having an excess of bits in the $Z$ basis becomes wasteful.   The choice of $P^B_X$ is governed by whether we want to design a system to function over wide ranges of loss as possible.  And if so, how great a decrease in secure key length are we willing to accept to achieve this robustness.

We've allowed $P^A_X$ to vary and in particular, we do not assume that $P^A_X=P^B_X$.  Some optimized values for $P^A_X$, corresponding to figure \ref{fig:fixed1} are presented in Appendix C.  In it is observed that $P^A_X$ is generally not equal to $P^B_X$, which means there is an advantage to allowing $P^A_X\ne P^B_X$.  Table \ref{tab:pb30} corresponds to $P^B_X=0.3$ and $p_{ec}=5\times 10^{-5}$, it shows that $P^A_X$ is close to one for $\eta_{\mathrm{loss}}$=18 dB, and decreases as the loss increases.  
This demonstrates that setting $P^B_X=0.3$ is sacrificing too much data to parameter estimation.  The opposite case is shown in table \ref{tab:pb90}, which corresponds to $P^B_X=0.9$ and $p_{ec}=10^{-6}$.  Here we see that for $\eta_{\mathrm{loss}}\ge 34$ dB, $P^A_X$ is less than $P^B_X=0.9$.  This occurs because the loss is so great that we need to ensure we have sufficient data in the $Z$-basis for parameter estimation. 

We've shown that allowing $P^A_X$ to differ from $P^B_X$ is beneficial when $P^B_X$ is fixed.  However, care must be taken with this observation.  We have not shown that one should always optimize {\it both} $P^A_X$ and $P^B_X$ separately.  Instead, we have argued that if $P^B_X$ is fixed, then it is advantageous to allow $P^A_X$ to differ from $P^B_X$.  In fact, one can prove that for any values of $P^A_X$ and $P^B_X$, one can find different values, where $P_X=P^A_X=P^B_X$, that yields the same ratio of raw bits in each basis and provides a possibly greater total number of sifted bits.  
The proof of this is presented in appendix D.

\begin{figure} 
\begin{center}
\begin{tabular}{c} 
\includegraphics[width=0.7\textwidth]{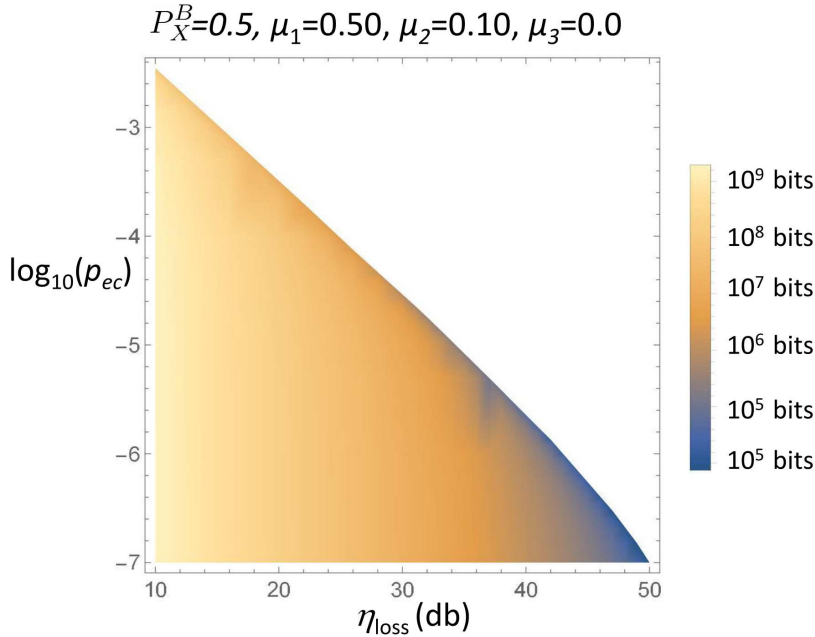}
\end{tabular}

   \caption[example] 
   { \label{fig:fixed2} 
A 3D plot of $\ell$ plotted for total system loss $\eta_{\mathrm{loss}}$ and $\log_{10}(p_{ec})$.  $QBER_I=0.01$, integration time was 30 minutes, $P^B_X=0.5$ was fixed, $\mu_1=0.50$, $\mu_2=0.10$ and $\mu_3=0$.  Even when both $P^B_X$ and $\{\mu_1,\mu_2,\mu_3\}$ are fixed, we can still choose values that allow one to extract a secure key over a wide range of system parameters.}
\end{center}
   \end{figure}

Due to deployment in different locations and operation during both night and day, the environmental factors of the system will differ. 
So far we assumed that signal intensities could be adjusted to accommodate the anticipated operating conditions. In practice this could be both costly and difficult to automatically employ, hence we now investigate the effect of fixing  $\{\mu_1,\mu_2,\mu_3\}$. We look at finding fixed values for the intensities that still allow $\ell>0$ over a wide range of different environmental conditions. As discussed in section 3.4, we use the numerically optimized results for $\ell$, with $P^B_X$ fixed, to help find robust values for $\mu_1$ and $\mu_2$, as in figure \ref{fig:fixed2} where $\ell$ is plotted  for $P^B_X$=0.5 and $\mu_1=0.50$, $\mu_2=0.10$ and $\mu_3=0$. Even when both $P^B_X$ and the intensities are fixed, we can still obtain $\ell>0$ over a wide ranges of values for $\eta_{\mathrm{loss}}$ and $p_{ec}$.  Furthermore, we obtain significant values for the secure key length. 
The value $P^B_X=0.5$ is not special and similar results can be obtained for other commonly available splitting ratios of beamsplitters (appendix B figure \ref{fig:D_fig2}).

\section{Effects of intensity uncertainty on secure key length}
\label{sec:uncertainty}
In the protocol the pulse intensities need to be well characterized.  In a laboratory setting the intensities of the WCPs can be controlled to a high degree.  In contrast, this will be difficult in a commercial system deployed within a busy urban environment.  Instead, there may be long terms drifts or else systematic offsets from the ideal intensity values.  This uncertainty affects the protocol as the security analysis assumes Alice transmits particular values for $\{\mu_j\}$.   Bob then uses this information to estimate the vacuum and single photon yields.  To ensure security, we must conservatively assume that the values transmitted correspond to those leading to the the largest amount of privacy amplification required.  

If the WCP intensities can vary by at most a fraction $f$, the actual mean photon number of the $j$-th intensity lies within the range $[\mu_j-f\mu_j,\mu_j+f\mu_j]$.  If we neglect the other intensities and model the variation from $\mu_j$ by a Gaussian distribution, then we would require that a large variation from $\mu_j$ must occur with probability less than $\epsilon_{sec}$, otherwise the security might be compromised. For the full protocol, we use 3 different WCPs, the weakest pulse the vacuum ($\mu_3=0$). Any variation in this will be due to dark counts and stray-light and this is already taken into account in the previous analysis. Instead, we are concerned with the uncertainty in $\mu_1$ and $\mu_2$. If we model these with a multivariable Gaussian, then we require that any fractional variation larger than $f$ occurs with probability less than $\epsilon_{sec}$.  The variance of these Gaussians is distinct from the photon statistical fluctuations but is determined by the uncertainty in how well one can set the intensities of the WCPs, an extremely stable laser could still have a large uncertainty in the WCPs intensities due to calibration of the electronic control system.

In practice, we can measure and determine the uncertainties in the WCPs and then deduce the value for the maximum  fractional deviation, $f$. Any larger deviations must occurs with probability less than $\epsilon_{sec}$.  From this we obtain a range of values for $\mu_1$ and $\mu_2$.  We find combinations of values that lead, in the worst case, to the smallest secure key.  As we have no a priori information about what are the real values for $\mu_1$ and $\mu_2$, we assume the worst case and perform parameter estimation and privacy amplification with those values of intensities.  This ensures the security of the system, but at a cost of reducing the length of the secret key.

Similar approaches have been studied before, but with important differences~\cite{rice2009, intensity2016}.  In~\cite{rice2009}, they employed a numerical parameter estimation method with possible values of $\mu_j$ whose uncertainties were accounted within the numerical optimization scheme. In~\cite{intensity2016} the authors use an analytic approach directly incorporating fluctuations within the estimation procedure. Here, we use an analytic estimation procedure but take account of fluctuations using numerical optimisation, also considering the case where the nominal intensities are fixed for different losses, not considered in~\cite{rice2009,intensity2016}. Additionally, in~\cite{rice2009,intensity2016} the same uncertainties are applied to all signals with the same intensity but different polarization, we employ a slightly more involved approach as described below. As such, our results complement the existing literature.

Separate lasers and driving electronics may be used for different signal states, hence we must allow $\mu_1$ and $\mu_2$ to vary independently for both bases and bit values, i.e. they can take different values from the intervals $[\mu_j-f\mu_j,\mu_j+f\mu_j]$, $j=1,2$.  To estimate $s_{X,0}$ and $s_{X,1}$ Bob will need to know $\mu_1$ and $\mu_2$.  We assume Bob knows both the intended values and $f$.  As such, he can construct the intervals $[\mu_j-f\mu_j,\mu_j+f\mu_j]$, $j$=1,2.  However, neither Bob or Alice will know the real values for $\mu_1$ and $\mu_2$.  During parameter estimation, Bob must choose values for $\mu_1$ and $\mu_2$ that lead to the smallest secret key.  To summarize, we choose values for $\mu_j$ from $[\mu_j-f\mu_j,\mu_j+f\mu_j]$, $j$=1 and 2, independently for the four signal states $H$, $V$, $D$ and $A$, and also independently for Bob.  
We then find the smallest possible value for $\ell$, which is the worst case value.  To guarantee that we have performed sufficient privacy amplification, and thus the protocol is secure, we must use the worst case value for the secure key length. 

To find the minimum value of the secret key, we discretise the intervals $[\mu_j-f\mu_j,\mu_j+f\mu_j]$. Numerical investigations showed that the minimum always occurred for values of $\mu_j$ at the edges of the intervals.  This lead us to limit ourselves to 3 possible values for each $\mu_j$: $\mu_j-f\mu_j$, $\mu_j$ and $\mu_j+f\mu_j$, retaining the middle value as a check on our assumption that the end values always yield the minimum.  The minimization thus requires evaluating the secure key length for $3^{10}$ different sets of WCP intensities.

\begin{figure} 
\begin{center}
\begin{tabular}{c} 
\includegraphics[width=\textwidth]{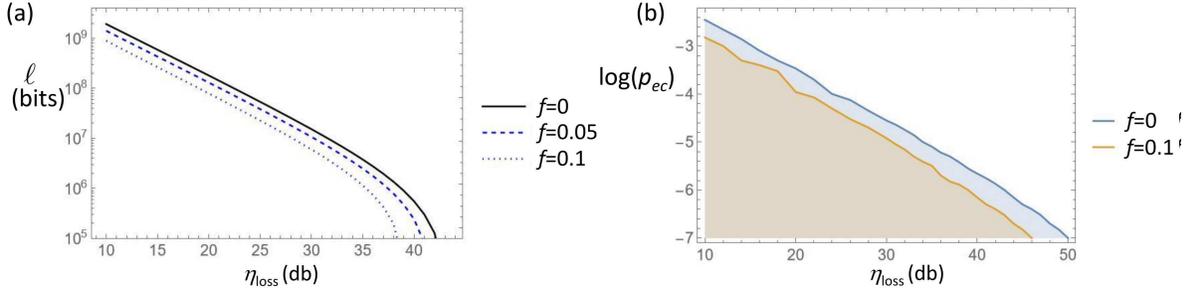}
\end{tabular}
\end{center}
   \caption[example] 
   { \label{fig:uncertainties} 
Effects on $\ell$ due to uncertainties in the WCP intensities. Integration time of 30 minutes, $QBER_I=0.01$, $P^B_X=0.5$, $\mu_1=0.5$, $\mu_2=0.1$ and $\mu_3=0$.  (a) $\ell$ against total system loss $\eta_{\mathrm{loss}}$ for different values of the fractional uncertainty, $f$, with $p_{ec}=10^{-6}$.  The black line corresponds to $f=0$, i.e. no uncertainty, while the dashed lines is for $f=0.05$ and the dotted lines are for $f=0.1$.   (b) Regions where $\ell>0$. $x$-axis is the total system loss $\eta_{\mathrm{loss}}$ and the $y$-axis is $\log_{10}(p_{ec})$.} 
\end{figure}

We study the case where we fix both the receiver basis probability and the WCP intensities, illustrated in figure \ref{fig:uncertainties}.  As expected, the effect of uncertainty is to reduce $\ell$ and in particular, $\ell$ goes to zero for lower values of $\eta_{\mathrm{loss}}$ as $f$ increases. The effect, for a given value of $p_{ec}$, is to decrease the total loss budget of about 4 dB which can make a significant difference and must be factored in when designing a QKD system.

\newpage

\section{Conclusions}
\label{sec:conc}
We have presented an in-depth investigation of the performance of efficient BB84 with the aim of aiding the design of terrestrial free-space QKD systems. For fibre-based systems, the channel conditions are relatively stable and controled, in contrast to a ground based, free-space QKD network where daily light levels and environmental conditions can vary site-to-site and from hour-to-hour. This motivates a systematic study of how the secure key length varies with loss and extraneous count rates. For the design of realistic commercial systems, for example see~\cite{airqkdref}, one must also consider fixing protocol parameters though operating in different environments. We examine such factors that are important to the deployment of a ground based QKD network.  


For a real system, the signals are transmitted for a finite duration, and thus finite statistical effects are important to the performance.  A particularly dramatic example was illustrated in figure \ref{fig:time}, where increasing the integration time allowed extraction of a secure key in conditions where this was impossible for smaller integration times.  Using a long integration times thus makes a system more robust.

For longer integration times, the secure key length was strongly affected by the probability of an extraneous count, e.g. from background light or a detector dark count.  In particular, we found that when the rate of background counts is high, we can tolerate only low losses if we are to generate a secret key.  This is an important design consideration.  To allow for as large a loss budget as possible, we must reduce the extraneous counts.  For a ground based, free-space QKD system, one should either operated in night conditions, or the receivers must be shielded from background light.

For a commercial QKD system, it would be difficult and expensive to customise all system parameters when deploying the system in different conditions.  For example, 
the receiver basis choice is usually implemented with a passive beam-splitter, which will be fixed during production to reduce costs.  
Similarly, changing the WCP intensities for different locations or channel conditions would require adjustable components and systems, potentially increasing the cost. Adapting the intensities either on a site-by-site basis, or on a session basis adds considerable complication to deployment and operations, further increasing costs.  
In section \ref{sec:fixed} we showed that one can fix {\it both} the receiver's basis choice probability and the WCP intensities, and still extract a secure key over a wide range of different environmental conditions.  We also found that a larger secure key is obtained by allowing the transmitters basis choice probability to differ from the receiver's, when the latter was fixed. 


Finally, we investigated the effects of uncertainties in the fixed WCP intensities, where we allow for independent uncertainties in all four encoded signal states.  If their is any uncertainties, then to ensure security, this needs to be factored in during the privacy amplification stage.  We find that uncertainties reduce the secure key length.  In particular, it can reduce the region of parameter space for which we can extract a secure key.  In some cases the reduction can be of the order of 3-4 dB.  As such it is vital to factor in these effects when designing any QKD system.  

\section*{Acknowledgement}
We acknowledge support from Innovate UK project AirQKD (Project number: 45364).  D. K. L. Oi acknowledges support from the UK NQTP, the Quantum Technology Hub in Quantum Communications (EPSRC Grant Ref: EP/T001011/1), and the EPSRC Researchers in Residence programme at the Satellite Applications Catapult (EPSRC Grant Ref: EP/T517288/1). The authors thank Gerald Bonner, Ross Donaldson, Yoann Noblet and Mike Parker for helpful discussions on the implementation of ground based free-space QKD.  We also thank Jasminder S. Sidhu, Duncan McArthur and Roberto G. Pousa for important discussion on simulating and optimizing efficient BB84 with WCPs.

\section*{Author contributions}
TB developed code and contributed to modelling, discussions, and interpretations of the research. All authors contributed to writing the manuscript. DO developed the idea for the work and contributed to modelling, discussions, and interpretations of the research.

\section*{Data availability}
All data is available on request from authors. SatQuMA code is released under the MIT Licence and is available at https://github.com/cnqo-qcomms/SatQuMA

\subsection*{ORCID iDs}
Thomas Brougham https://orcid.org/0000-0002-9066-1771\\ 
Daniel K. L. Oi  https://orcid.org/0000-0003-0965-9509

\section*{Appendix A: Examples of calculating loss budgets for QKD network design}
The results presented in sections 4 and 5 can be used to design optical systems needed to realize a ground based, free-space QKD.  In particular, we can calculate the total loss budget allocated for the free-space optics. We illustrate this by some simple examples.  

Suppose we plan to setup a QKD link in an area that is prone to fog.  The background light levels are measured at various times and used to estimate $p_{ec}$.  
At the site, the distance between the transmitter and receiver is 300m; and the quantum source operates at 850nm.  Using Modtran 6 \cite{modref} we can calculate the losses due to scattering and absorption for this distance, during different weather conditions.  The site is prone to advection fog with a visibility of 800m, which at 300m gives a loss of $\approx 6.4$ dB from scattering and absorption.     

If we can adapt our choice of both $P^B_X$ and WCP intensities specifically for the site, then we can use the results presented in figure \ref{fig:min30}.  For $p_{ec}=10^{-4}$ and $QBER_I=0.01$, $\ell>0$ for values of $\eta_{\mathrm{loss}}$ up to and including $\eta_{\mathrm{loss}}=22$ dB.  Losses from sources other than scattering and absorption, such as from detectors, internal optics, diffraction, turbulence and pointing errors, must be be no greater than 15.6 dB.  The transmitter and receiver must thus be designed so as to keep these losses below about 15 dB.  However, if the extraneous counts can be reduced by a factor of 4, to $p_{ec}=2.5\times 10^{-5}$, by better design of the transmitter and receiver, then we can extract a key up to $\eta_{\mathrm{loss}}=26$ dB. This increases the loss budget for the design of the optics to 19.6 dB.  This example illustrates that if the QKD link is to be robust to advection fog, then it is crucial to design the receiver and transmitter so as to reduce background light.

In the previous example we assumed that $P^B_X$ and the WCP intensities can both be adapted to the location.  In a future commercially deployed systems, such as that planned as part of the AirQKD project \cite{airqkdref}, both the receiver basis and the set of intensities will be fixed for all units.  Furthermore, suppose the design brief requires each link to produce 50, 256-bit AES keys every 10 minutes.  For an integration time of 30 minutes, we need $\ell=38,400$ bits.  As with the previous example, we assume that the link distance is 300m and the source is at 850nm.  Furthermore, locations are chosen such that for night operation, $p_{ec}$ is at most $10^{-6}$.  Using MODTRAN 6 \cite{modref}, for an urban environment with link distance 300m, the losses due to scattering and absorption are 0.6 dB.  We fix the protocol parameters to $P^B_X=0.5$, $\mu_1=0.5$, $\mu_2=0.1$ and $\mu_3=0$, which are the same as for plot \ref{fig:fixed2}.  By assumption, the the worst case is $p_{ec}=10^{-6}$, for which $\ell> 38,400$ for $\eta_{\mathrm{loss}}$ up to and including 42 dB.  The total loss budget, minus absorption and scattering, is approximately 41 dB.

For $\eta_{\mathrm{loss}}=42$ dB, we produce over 125,800 bits of key.  However, due to finite statistical effects, the key dips to below 38,400 bits for $\eta_{\mathrm{loss}}=42.5$ dB.  If we operate the system right at the edge of the loss budget, then any unexpected increase in say pointing error due to strong winds, will lead to the system falling below its aim of producing 50 AES keys every 10 minutes.  However, as we produce large amounts of key (125,800 bits for $\eta_{\mathrm{loss}}$), we could store the excesses key.  It is thus possible to smooth out any outages by designing the memory and the classical part of the cryptographic infrastructure, using knowledge of the protocol performance under different environmental conditions. 


\section*{Appendix B: Additional result for different values of $P^B_X$}
\begin{figure} 
\begin{center}
\begin{tabular}{c} 
\includegraphics[width=\textwidth]{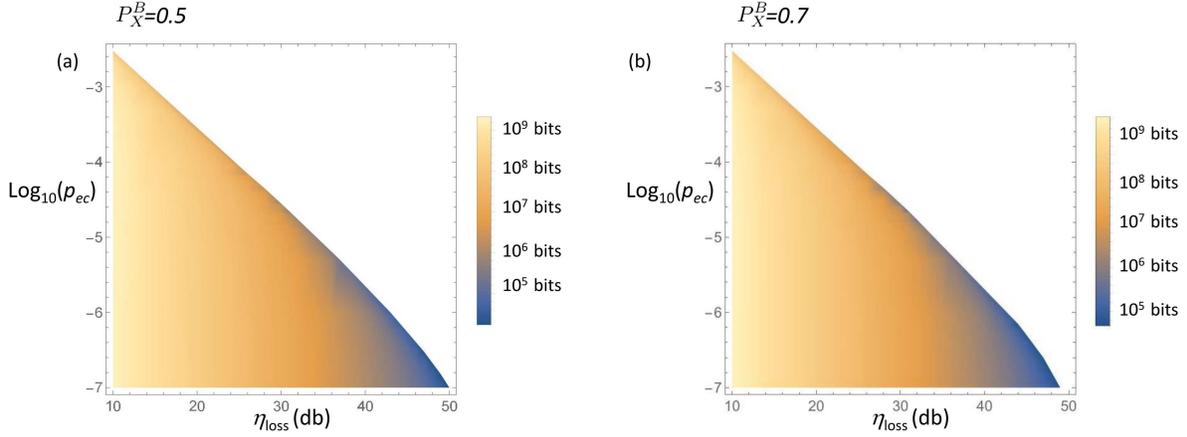}
\end{tabular}
\end{center}
   \caption[example] 
   { \label{fig:D_fig1} 
A 3D plot of $\ell$ plotted for total system loss $\eta_{\mathrm{loss}}$ and $\log_{10}(p_{ec})$.  All plots are for $QBER_I=0.01$ and an integration time of 30 minutes.  The receiver basis choice probability is fixed to set values for both plots.  In figure (a) $P^B_X=0.5$ and in (b) $P^B_X=0.7$.}
   \end{figure}

In section 5 we investigated fixing the receiver basis choice probability, $P^B_X$.  We demonstrated that $P^B_X$ can be fixed to either 0.3 and 0.9, and $\ell>0$ for a wide range of values for $p_{ec}$ and $\eta_{\mathrm{loss}}$.  The two values considered for $P^B_X$ correspond to common commercially available beam-splitter splitting ratios.  Two other common values are $P^B_X=0.5$ and $0.7$, which we now investigate for completeness.  
Figure \ref{fig:D_fig1} is a 3D plot of $\ell$, plotted for $\eta_{\mathrm{loss}}$ and $\log_{10}(p_{ec})$.  Both plots are for an integration time of 30 minutes and $QBER_I=0.01$.  In figure \ref{fig:D_fig1}, (a) is for $P^B_X=0.5$, while (b) is for $P^B_X=0.7$.  This plot demonstrates that one can obtain a secure key for a wide range of environmental conditions for many different fixed values of $P^B_X$.

\begin{figure} 
\begin{center}
\begin{tabular}{c} 
\includegraphics[width=\textwidth]{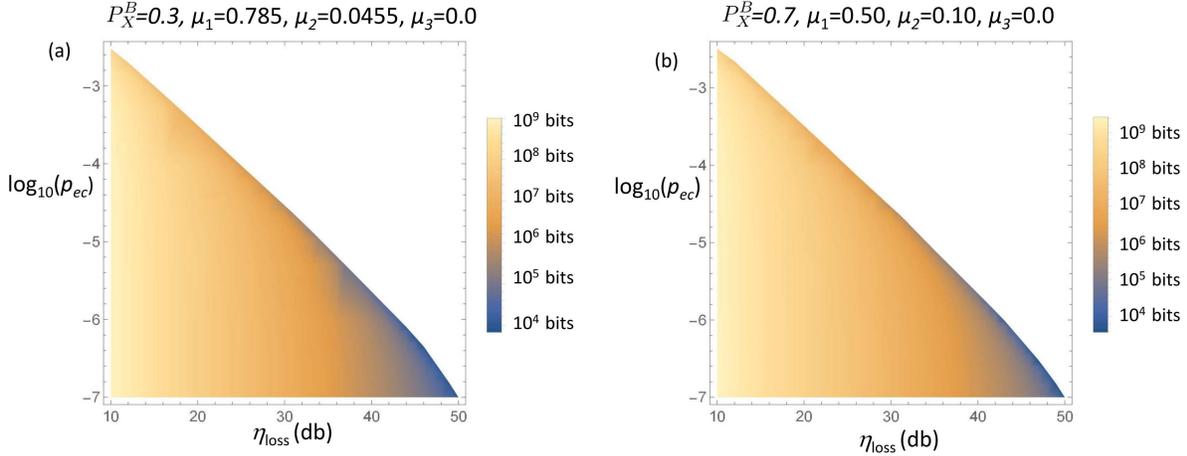}
\end{tabular}
\end{center}
   \caption[example] 
   { \label{fig:D_fig2} 
A 3D plot of $\ell$ plotted for total system loss $\eta_{\mathrm{loss}}$ and $\log_{10}(p_{ec})$.  All plots are for $QBER_I=0.01$ and an integration time of 30 minutes.  Both the receiver basis choice probability and the source intensities are fixed to set values for both plots.  In figure (a) $P^B_X=0.3$, $\mu_1=0.785$, $\mu_2=0.0455$ and $\mu_3=0$, while in (b) $P^B_X=0.7$, $\mu_1=0.50$, $\mu_2=0.10$ and $\mu_3=0$.}
\end{figure}

We also investigated fixing both $P^B_X$ and $\{\mu_1,\mu_2,\mu_3\}$. Even when fixing all these parameters, it was still possible to find values such that one could obtain secure key over a wide range of environmental conditions, as shown for $P^B_X=0.5$ in figure \ref{fig:fixed2}.  We can also obtain similar results for different values of $P^B_X$.  In figure \ref{fig:D_fig2} we plot $\ell$ against $\log_{10}(p_{ec})$ and $\eta_{\mathrm{loss}}$, for an integration time of 30 minutes and $QBER_I=0.01$.  In plot (a) $P^B_X=0.3$, $\mu_1=0.785$, $\mu_2=0.0455$ and $\mu_3=0$, while in (b) $P^B_X=0.7$, $\mu_1=0.50$, $\mu_2=0.10$ and $\mu_3=0$.  The results demonstrate that one can obtain fixed intensities for different choice of $P^B_X$, which lead to large values of the secure key over wide ranges of environmental parameters.

\section*{Appendix C: Sample optimized values for $P^A_X$, when $P^B_X$ is fixed}
In this appendix we present some optimized values for $P^A_X$ when $P^B_X$ is fixed.  The fact that we allow Alice and Bob to have different basis probabilities is a novel feature of the current analysis.  As such, it is worth presenting some results for how $P^A_X$ changes.  The results are presented in tables 2-5.  All results correspond to an integration time of 30 minutes and $QBER_I=0.01$.

\begin{table}[h!]
\caption{A table showing the optimalized values for $P^A_X$, when $P^B_X=0.3$.  All values are for $p_{ec}=5\times 10^{-5}$.} 
\label{tab:pb30}
\begin{center}       
\begin{tabular}{|l|l|l|}
\hline
\rule[-1ex]{0pt}{3.5ex}  $\eta_{\mathrm{loss}}$ (dB) & $P^B_X$ & $P^A_X$\\
\hline
\rule[-1ex]{0pt}{3.5ex}  18 & 0.3 & 0.97964\\
\hline
\rule[-1ex]{0pt}{3.5ex}  20 & 0.3 & 0.96525\\
\hline
\rule[-1ex]{0pt}{3.5ex}  22 & 0.3 & 0.95672\\
\hline
\rule[-1ex]{0pt}{3.5ex}  24 & 0.3 & 0.94319 \\
\hline
\rule[-1ex]{0pt}{3.5ex}  26 & 0.3 & 0.91283\\
\hline
\end{tabular}
\end{center}
\end{table}

\begin{table}[h!]
\caption{A table showing the optimalized values for $P^A_X$, when $P^B_X=0.5$.  All values are for $p_{ec}=5\times 10^{-5}$.} 
\label{tab:pb50}
\begin{center}       
\begin{tabular}{|l|l|l|}
\hline
\rule[-1ex]{0pt}{3.5ex}  $\eta_{\mathrm{loss}}$ (dB) & $P^B_X$ & $P^A_X$\\
\hline
\rule[-1ex]{0pt}{3.5ex}  18 & 0.5 & 0.95482\\
\hline
\rule[-1ex]{0pt}{3.5ex}  20 & 0.5 & 0.96045\\
\hline
\rule[-1ex]{0pt}{3.5ex}  22 & 0.5 & 0.95178\\
\hline
\rule[-1ex]{0pt}{3.5ex}  24 & 0.5 & 0.94165 \\
\hline
\rule[-1ex]{0pt}{3.5ex}  26 & 0.5 & 0.89656\\
\hline
\end{tabular}
\end{center}
\end{table}

\begin{table}[h!]
\caption{A table showing the optimalized values for $P^A_X$, when $P^B_X=0.7$.  All values are for $p_{ec}=10^{-6}$.} 
\label{tab:pb70}
\begin{center}       
\begin{tabular}{|l|l|l|}
\hline
\rule[-1ex]{0pt}{3.5ex}  $\eta_{\mathrm{loss}}$ (dB) & $P^B_X$ & $P^A_X$\\
\hline
\rule[-1ex]{0pt}{3.5ex}  32 & 0.7 & 0.92312\\
\hline
\rule[-1ex]{0pt}{3.5ex}  34 & 0.7 & 0.90846\\
\hline
\rule[-1ex]{0pt}{3.5ex}  36 & 0.7 & 0.87179\\
\hline
\rule[-1ex]{0pt}{3.5ex}  38 & 0.7 & 0.86605 \\
\hline
\rule[-1ex]{0pt}{3.5ex}  40 & 0.7 & 0.80577\\
\hline
\end{tabular}
\end{center}
\end{table}

\begin{table}[h!]
\caption{A table showing the optimalized values for $P^A_X$, when $P^B_X=0.9$.  All values are for $p_{ec}=10^{-6}$.} 
\label{tab:pb90}
\begin{center}       
\begin{tabular}{|l|l|l|}
\hline
\rule[-1ex]{0pt}{3.5ex}  $\eta_{\mathrm{loss}}$ (dB) & $P^B_X$ & $P^A_X$\\
\hline
\rule[-1ex]{0pt}{3.5ex}  32 & 0.9 & 0.91367\\
\hline
\rule[-1ex]{0pt}{3.5ex}  34 & 0.9 & 0.89787\\
\hline
\rule[-1ex]{0pt}{3.5ex}  36 & 0.9 & 0.86897\\
\hline
\rule[-1ex]{0pt}{3.5ex}  38 & 0.9 & 0.78494 \\
\hline
\rule[-1ex]{0pt}{3.5ex}  40 & 0.9 & 0.72154\\
\hline
\end{tabular}
\end{center}
\end{table}

A key feature in all tables is that the optimal secure key length is obtained for $P^A_X\ne P^B_X$.  As a general trend, the value of $P^A_X$ tends to decrease as $\eta_{\mathrm{loss}}$ increases.  Intuitively, this occurs as for greater values of $\eta_{\mathrm{loss}}$ we have less overall data and thus less data in the $Z$-basis.  To ensure we can perform parameter estimation without error, we need to increase the fraction of data in the $Z$-basis.

\section*{Appendix D: Proof that setting $P^A_X=P^B_X$ yields the optimal raw key length}
In section \ref{sec:fixed} we argued that if $P^B_X$ is fixed, then it is advantageous to allow $P^A_X\ne P^B_X$ and to optimize Alice's basis probability choice independently.  Naively, this might be taken to suggest that we should always optimize both $P^A_X$ and $P^B_X$ independently.  This, however, is not necessarily true.  When {\it both} Alice and Bob's basis choices are free to vary, then it is sufficient to optimize $P_X=P^A_X=P^B_X$.  This follows from the following observation. For any values for $P^A_X$ and $P^B_X$, there exists a basis probability $P_X$ that yields the same ratio of raw bits in each basis, but where the total sifted bits is greater than or equal to number of raw bits obtained with $P^A_X$ and $P^B_X$.  This means that the optimal raw key can always be found by using $P^A_X=P^B_X =P_X$.

Suppose we have particular values for $P^A_X$ and $P^B_X$, where $P^A_X\ne P^B_X$. The fraction of bits retained after basis sifting is
\begin{equation}
\label{appendix1}
F=P^A_X P^B_X +(1-P^A_X)(1-P^B_X).
\end{equation}
The fraction of sifted bits in the $X$ basis relative to the $Z$ basis is $K=P^A_X P^B_X/[(1-P^A_X)(1-P^B_X)]$. Consider instead if Alice and Bob both had both chosen the $X$-basis with probability $P_X$.  The fraction of sifted bits retained is $F'=P^2_X +(1-P_X)^2$.  
For the ratio of bits in each basis to be the same as before we need $P_X/(1-P_X) =\pm \sqrt{K}$.  This yields two values for $P_X$, we take the value  $P_X=\sqrt{K}/(1+\sqrt{K})$.  One can show that 
\begin{equation}
\label{appendix2}
F'=\frac{1+K}{(1+\sqrt{K})^2}.
\end{equation}
Eq. (\ref{appendix1}) can be re-written in terms of $K$ as: $F=(1-P^A_X)(1-P^B_X)[1+K]$.  Our proposition is false if there exists values for $P^A_X$ and $P^B_X$ such that $F>F'$.  Using Eq. (\ref{appendix2}) and re-arranging, the condition $F>F'$ can be shown to be equivalent to 
\begin{equation}
\label{appendix3}
\sqrt{P^A_X P^B_X}+\sqrt{(1-P^A_X)(1-P^B_X)}>1.
\end{equation}
The right hand side of the above Eq. is of the form of a scalar product between the two vectors: $(\sqrt{P^A_X},\sqrt{1-P^A_X})$ and $(\sqrt{P^B_X},\sqrt{1-P^B_X})$.  From the Cauchy-Schwartz inequality, we have $(\sqrt{P^A_X},\sqrt{1-P^A_X})\cdot (\sqrt{P^B_X},\sqrt{1-P^B_X})\le [P^A_X+1-P^A_X][P^B_X+1-P^B_X]=1$.  This implies that the inequality (\ref{appendix3}) can never be satisfied.  But this means that the choice of $P_X$ that leads to Eq. (\ref{appendix2}) can never yields a smaller total fraction of sifted bits than for $P^A_X\ne P^B_X$, i.e. $F'\ge F$.  The raw key is thus greater than or equal to the original raw key.

\section*{Appendix E: Details for how $p_{ec}$ and $QBER_I$ contribute to the secure key length}

The secure key length, $\ell$, is calculated using Eqs. (\ref{skl}), (\ref{finitecor}) and (\ref{finitecor2}), together with Eqs. (1) to (5) from \cite{Lim2014}.  The extraneous count probability, $p_{ec}$, and the intrinsic quantum bit error rate, $QBER_I$, induce the errors, which in turn affects $\ell$.  We detail the relationship in this appendix.  The approach is the same as outlined in \cite{Sidhu2021}.

The average probability for a pulse of intensity $k\in\{\mu_1,\mu_2,\mu_3\}$ to produce a detection event is given by the equation 
\begin{equation}
    D_k =(1+p_{ap})\left[1-(1-2p_{ec})\right]\exp(-\eta_{\mathrm{loss}} k).
\end{equation}
The probability for a pulse of intensity $k$ to yield a bit with an error is given by
\begin{equation}
    e_k=p_{ec}+\frac{p_{ap}D_k}{2}+QBER_I (1-\exp[-\eta_{\mathrm{loss}} k]).
\end{equation}
The total number of pulses transmitted is the product of the integration time, $\tau$, and the source's repetition rate, $f_s$.  

If the channels parameter, such as $\eta_{\mathrm{loss}}$, can vary with time, then we partition the integration time into time-slots of width $\Delta\tau$, where this is chosen such that the parameters are approximately constant over this time interval. The number of detection events in the $X$-basis within the $j$-th time-slot are:
\begin{equation}
    n^{(j)}_{X,k}=P^A_X P^B_X p_k D_k (j) f_s \Delta\tau,
\end{equation}
where $p_k$ is the probability to transmit a pulse with intensity $k$.  The total number of detection events in the $X$-basis from pulses with intensity $k$ is: $n_{X,k}=\sum_j  n^{(j)}_{X,k}$ and the total number of detection events in the $X$-basis is $n_{X}=\sum_{k}  n_{X,k}$. The results for the $Z$-basis have the same form, but with $X$ swapped to $Z$.

The number of errors in the $X$-basis, within the $j$-th time-slot, is
\begin{equation}
    m^{(j)}_{X}=\frac{\sum_k {p_k e_k (j)}n^{(j)}_{X} }{\sum_k p_k D_k (j)},
\end{equation}
where $n^{(j)}_{X}=\sum_{k} n^{(j)}_{X,k}$. The total number of errors in the $X$-basis is $m_X=\sum_j m^{(j)}_{X}$.  The fraction of errors resulting from a pulse of intensity $k$ can be found by weighting $m^{(j)}_{X}$ by the term $p_k D_k (j)/[\sum_k p_k D_k (j)]$.  Similar results can be found for the $Z$ basis.  These quantities can be used in Eq. (\ref{finitecor}) and (\ref{finitecor2}) together with Eqs. (1) to (5) of \cite{Lim2014}, to calculate the length of the secure key (\ref{skl}).


\end{document}